\documentclass[%
aps,
prl,
twocolumn,
superscriptaddress,
amsmath,
amssymb,
citeautoscript
]{revtex4-1}
\usepackage{graphicx}
\usepackage{dcolumn}
\usepackage{bm}
\usepackage{hyperref}

\usepackage{xcolor}
\usepackage{soul}
\usepackage[normalem]{ulem}

\newcommand{\RNum}[1]{\uppercase\expandafter{\romannumeral #1\relax}}

\newcommand{\sitransition}{S1}
\newcommand{\siperimeter}{S2}
\newcommand{\sisolid}{S3}
\newcommand{\sisameactivity}{S4}
\newcommand{\sipolar}{S5}
\newcommand{\sipair}{S6}
\newcommand{\siinitial}{S7}
\newcommand{\siarea}{S8}
\newcommand{\sidispersion}{S9}

\newcommand{\sectionnematic}{\RNum{1}}
\newcommand{\sectiontransition}{\RNum{2}}
\newcommand{\sectionpolar}{\RNum{3}}
\newcommand{\sectionpair}{\RNum{4}}
\newcommand{\sectionarea}{\RNum{5}}
\newcommand{\sectiondispersion}{\RNum{6}}

\newcommand{\moviesorting}{S1}

\newcommand{\zc}{\zeta_{\rm{c}}}
\newcommand{\ze}{\zeta_{\rm{e}}}

\begin{document}

\preprint{APS/123-QED}

\title{
Cell Sorting in an Active Nematic Vertex Model
}

\author{Jan Rozman}
\email{jan.rozman@physics.ox.ac.uk}
\affiliation{Rudolf Peierls Centre for Theoretical Physics, University of Oxford, Oxford OX1 3PU, United Kingdom}

\author{Julia M. Yeomans}%
\affiliation{Rudolf Peierls Centre for Theoretical Physics, University of Oxford, Oxford OX1 3PU, United Kingdom}

\date{\today}

\begin{abstract}
We study a mixture of extensile and contractile cells using a vertex model extended to include active nematic stresses. The two cell populations phase separate over time. While phase separation strengthens monotonically with an increasing magnitude of  contractile activity, the dependence on extensile activity is non-monotonic, so that sufficiently high values reduce the extent of sorting. We interpret this by showing that extensile activity renders the system motile, enabling cells to undergo neighbour exchanges. Contractile cells that come into contact as a result are then more likely to stay connected due to an effective attraction arising from contractile activity.
\end{abstract}

\maketitle

{\textit{Introduction}}---Phase separation in biological systems plays a role on scales ranging from mixtures of RNA and charged proteins~\cite{dutagaci2021charge} to populations of cells~\cite{krens2011cell}. In particular, cell sorting has been observed, e.g., in \textit{in vitro} mixtures of cell from different tissues~\cite{moscona1952dissociation,townes1955directed} and hydra regeneration~\cite{gierer1972regeneration,cochet2017physical}. It is suggested to underpin the formation of boundaries between regions of different cells in developmental contexts~\cite{krens2011cell,batlle2012molecular} and has been implicated in the organisation of tumors~\cite{pawlizak2015testing}. Therefore, the mechanisms behind it have long been of interest. In an important step Steinberg proposed the Differential Adhesion Hypothesis to explain cell sorting, suggesting that it arises from differences in adhesion between like and unlike cell types~\cite{steinberg1963reconstruction,foty2005differential}. However, biological systems are active~\cite{ramaswamy2010mechanics,marchetti2013hydrodynamics}, allowing for sorting not driven by equilibrium energy minimisation. The most well known active phase separation mechanism is the motility-induced phase separation ~\cite{tailleur2008statistical,cates2015motility,gonnella2015motility} of self-propelled particles forming a dense and a dilute phase.

Active nematics are a type of active matter characterised by components generating dipolar stresses, resulting in chaotic flows termed active turbulence~\cite{simha2002hydrodynamic,doostmohammadi2018active}. They have been used to model, e.g., Madin-Darby canine kidney (MDCK) cells, a type of epithelium~\cite{saw2017topological},   fibroblasts~\cite{duclos2014perfect,duclos2017topological}, and bacterial colonies~\cite{dell2018growing}. Of particular relevance to this Letter, Ref.~\cite{balasubramaniam2021investigating} reports that wild type MDCK cells resemble extensile active nematics, whereas E-cadherin knock-out MDCK cells resemble contractile active nematics. In a mixture of wild type (extensile) and E-cadherin knock-out (contractile) MDCK cells, the two populations phase separate. Phase separation in mixtures of active nematics with different activities has also been demonstrated in theoretical models~\cite{balasubramaniam2021investigating,bhattacharyya2023phase,graham2024cell,bhattacharyya2024phase}. 

2D vertex models, which represent epithelial cells as  polygons tiling a surface, are a powerful tool for understanding tissue mechanics~\cite{honda1980much,farhadifar2007influence,fletcher2014vertex,alt2017vertex}. Examples of phase separation mechanisms in vertex models are the interplay of local alignment and contact inhibition of locomotion in polar models which produces phase separation in cell density~\cite{lin2018dynamic}, as well as  different amplitudes of fluctuating tensions on cell-cell junctions~\cite{krajnc2020solid} and, on small scales, differences in cell shape~\cite{sahu2020sorting} which  lead to cell sorting. Recent work has shown that it is possible to generalise vertex models to include active nematic driving~\cite{  comelles2021epithelial,duclut2022active,sonam2023mechanical,lin2023structure,rozman2023shape,rozman2023dry}. Therefore, given the relevance of nematic activity to tissue dynamics, we investigate cell sorting in an active nematic vertex model. We find that  a mixture of extensile and contractile cells (i.e. cells where the active stresses tend to either extend or to contract them along their long axis) phase separates over time, with two mechanisms contributing to the separation:  i) the motility of the extensile cells and ii) contractile activity causing an effective attraction between contractile cells.

\begin{figure*}
    \includegraphics{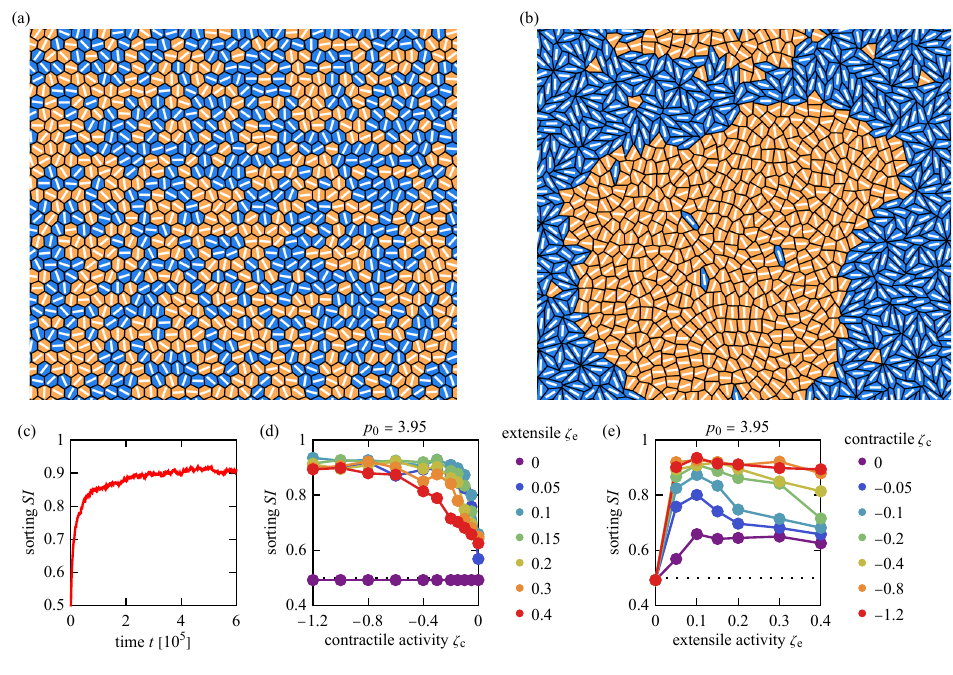}
    \caption{
    (a) Initial condition of the simulation: blue cells are extensile ($\ze=0.1$) and orange cells are contractile ($\zc=-0.4$). White lines show cell directors. (b) Model tissue at $t=6\times10^5$, showing phase separation between the extensile and contractile cells. (c) Segregation index $SI$ as a function of time for the simulation shown on panels (a),(b). (d),(e) Segregation index at $t=6\times10^5$ as a function of contractile activity at different extensile activities (d) and as a function of extensile activity at different contractile activities (e). Dotted lines on panels (d) and (e)  show $SI=0.5$. $k_\text{P}=0.02$ and $p_0=3.95$ for all panels.
    }
    \label{fig:sorting}
\end{figure*}

{\textit{The model}}---We use the area and perimeter elasticity vertex model~\cite{farhadifar2007influence} extended to include nematic activity~\cite{sonam2023mechanical,lin2023structure}. The dimensionless energy function is $e_{\rm VM}=\sum_{i}\left[ \frac{1}{2}\left(a_i - 1\right)^2 + \frac{k_\text{P}}{2}\left(p_i - p_0\right)^2\right]$. Here, the sum is over all $N_\text{c}$ cells, $a_{i}$ is the area of cell $i$, $k_\text{P}$ is the perimeter elasticity modulus, $p_{i}$ is the perimeter of cell $i$, and $p_0$ is the target perimeter. $p_0$ determines the mechanical properties of the passive model, controlling its rigidity transition at a critical value $p_*$, which was first reported at 3.81~\cite{bi2015density}. Further studies place the rigidity transition threshold value of the purely passive vertex model in the range $\approx 3.81-3.94$~\cite{merkel2019minimal,wang2020anisotropy,tong2022linear}. 

Dynamics are given by the overdamped equation of motion, which reads in dimensionless form: $\dot{\mathbf{r}}_{j} =  - \nabla_{\mathbf{r}_{j}} e_\text{VM} + \mathbf{f}_{j}^\text{act}$, where $\mathbf{r}_{j}$ is the position of vertex $j$, the overdot indicates the time derivative, $\nabla_{\mathbf{r}_{j}} $ is the gradient with respect to the position of vertex $j$, and $\mathbf{f}_{j}^\text{act}$ is the active force on vertex $j$. We define the active nematic forces following Ref.~\cite{lin2023structure}. Each cell $i$ experiences a bulk stress
\begin{equation}
    \boldsymbol{\sigma}_i=-\zeta_i \mathbf{Q}_i,
\end{equation}
where $\zeta_i$ is the activity of the cell and $\mathbf{Q}_i=\left(\frac{1}{p_i}\sum_{k} \ell_k\hat{\mathbf{t}}_k\otimes \hat{\mathbf{t}}_k\right)-\frac{1}{2}\mathbf{I}$ is a cell shape anisotropy tensor. The sum is along all junctions $k$ of cell $i$, $\ell_k$ is the length of junction $k$, $\hat{\mathbf{t}}_k$ a unit vector along the direction of the junction, and $\mathbf{I}$ is the identity tensor. The stress tensor is converted into vertex forces using the approach proposed by Tlili \textit{et al.}~\cite{tlili2019shaping,lin2022implementation} (see Supplemental Material,  Sec.~\sectionnematic~\cite{SI}). Positive values of $\zeta$ correspond to extensile activity, whereas negative values correspond to  contractile activity. In the case where all cells have the same extensile activity ($\zeta_i=\ze>0$) and the activity is above a threshold (which falls to $0$ for $p_0>p_*$), this model has been shown to produce chaotic flows resembling extensile active nematic turbulence~\cite{lin2023structure}. 

We solve the equations of motion using a simple Euler scheme with time step $\Delta t = 0.01$. If the length of a junction falls below a threshold $\ell_{\rm T1}=0.01$, a T1 transition is performed, with final junction length set to $\ell'_{\rm T1}=1.1\times \ell_{\rm T1}$ (Fig.~\sitransition; see Supplemental Material, Sec.~\sectiontransition~\cite{SI} for discussion on how $\ell_{\rm T1}$ and $\ell'_{\rm T1}$ affect simulation results). Unless otherwise specified, we set $k_\text{P} = 0.02$ and $p_0=3.95$, corresponding to the fluid phase of the passive vertex model. We use a system of 1024 cells with periodic boundary conditions. The initial configuration is a 32 by 32 cell hexagonal lattice, with all vertices displaced by a small random perturbation so that cells start with a finite $\mathbf{Q}_i$ tensor. Without the perturbation, all components of $\mathbf{Q}_i$ would be 0 for a regular hexagon, resulting in no active forces at the start of the simulation. Simulations are run until $t=6\times 10^5$ unless otherwise specified.\\


{\textit{Results}}---We begin by simulating a model tissue in which half of the cells are randomly selected to be extensile ($\zeta_i = \ze$) whereas the remainder are contractile ($\zeta_i = \zc$). The initial configuration is shown in Fig.~\ref{fig:sorting}(a). The extensile and contractile cells phase separate over time [Fig.~\ref{fig:sorting}(b) and Movie~\moviesorting]. To quantify the extent of phase separation, we use the segregation index $SI=\frac{1}{N_\text{c}}\sum_{i=1}^{N_\text{c}}\frac{n_i}{n_i+n_i'}$~\cite{skamrahl2023cellular,graham2024cell}, where $n_i$ is the number of neighbours of cell $i$ that are of the same type and $n_i'$ is the number of neighbours of a different type. Figure~\ref{fig:sorting}(c) shows how the $SI$ changes over time for $\ze=0.1$ and $\zc=-0.4$. While the speed of the sorting slows down over time, almost all extensile cells form a single contiguous region by the end of the simulation [Fig.~\ref{fig:sorting}(b)]. The same is also true for the contractile cells.

Scanning over extensile and contractile activities, we find that the $SI$ at the end of the simulation depends monotonically on $\zc$ so that a higher magnitude of contractile activity results in better (or equal) phase separation [Fig.~\ref{fig:sorting}(d)]. However, although extensile activity is required for phase separation, the $SI$ depends on it non-monotonically. Increasing $\ze$ above an optimal value reduces sorting [Fig.~\ref{fig:sorting}(e)]. 

The sorting remains similar at $p_0=3.85$, closer to the originally reported threshold of the passive vertex model rigidity transition $p_*=3.81$~\cite{bi2015density} (Fig.~\siperimeter). Moreover, the extensile-contractile mixture still separates if $p_0$ is below $p_*$ (Fig.~\sisolid~\cite{SI}): the dependence on contractile activity remains monotonic and the dependence on extensile activity non-monotonic. However, as shown in Ref.~\cite{lin2023structure}, extensile activity must be above a threshold value to fluidise the tissue if $p_0<p_*$, whereas any extensile activity should fluidise it above $p_*$ [see Fig. 3(e) in Ref.~\cite{lin2023structure} for the phase diagram]. Correspondingly, sorting only takes place if $\ze$ is sufficiently high, approximately matching the reported value for active tissue fluidisation. Moreover, an extensile-passive or extensile-extensile mixture also features some phase separation, whereas a contractile-passive or contractile-contractile one does not (Fig.~\sisameactivity~\cite{SI}).\\

\begin{figure}
    \includegraphics{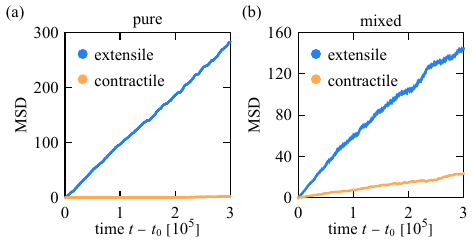}
    \caption{
    (a) Mean-squared displacement (MSD) of a system of only extensile ($\ze=0.1$) or only contractile ($\zc=-0.4$) cells. (b) MSD of a 50:50 mixture of extensile ($\ze=0.1$) and contractile ($\zc=-0.4$) cells. $k_\text{P}=0.02$ and $p_0=3.95$ for both panels.
    }
    \label{fig:msd}
\end{figure}

To understand the mechanisms behind phase separation, we first measure cell motility in the model using the mean-squared displacement of cells $\text{MSD}(t)=\frac{1}{N_\text{c}}\sum_{i=1}^{N_\text{c}}\left|\mathbf{\text{r}}^{(c)}_i(t)-\mathbf{\text{r}}^{(c)}_i(t_0)\right|^2$. Here, ${N_\text{c}}$ is the total cell number, $\mathbf{\text{r}}^{(c)}_i(t)$ is the position of the centre (mean of vertices) of cell $i$ at time $t$, and $t_0=3\times 10^5$ is the starting time for MSD measurements. Only extensile activity is reported to result in the system behaving as a motile active fluid  (provided activity is above a threshold if $p_0<p_*$)~\cite{lin2023structure}. As expected, cells in a purely extensile system move diffusively and are much more motile than cells in a purely contractile system [Fig.~\ref{fig:msd}(a)]. In a mixture, contractile cells become more motile, but still move less than their extensile counterparts [Fig.~\ref{fig:msd}(b)]. Phase separation due to differential diffusivity is a known mechanism~\cite{weber2016binary,mccarthy2024demixing} and it has been suggested to drive extensile-contractile phase separation in an active nematic phase-field model~\cite{graham2024cell}. However,  Ref.~\cite{mccarthy2024demixing} reports that differential diffusivity does not induce phase separation at high packing fractions in a particle-based  or a Voronoi model. The latter is closely related to the vertex model, and both represent confluent tissues, corresponding to packing fraction one. This suggests that other mechanisms are necessary to cause the cell sorting observed here.

\begin{figure*}
    \includegraphics{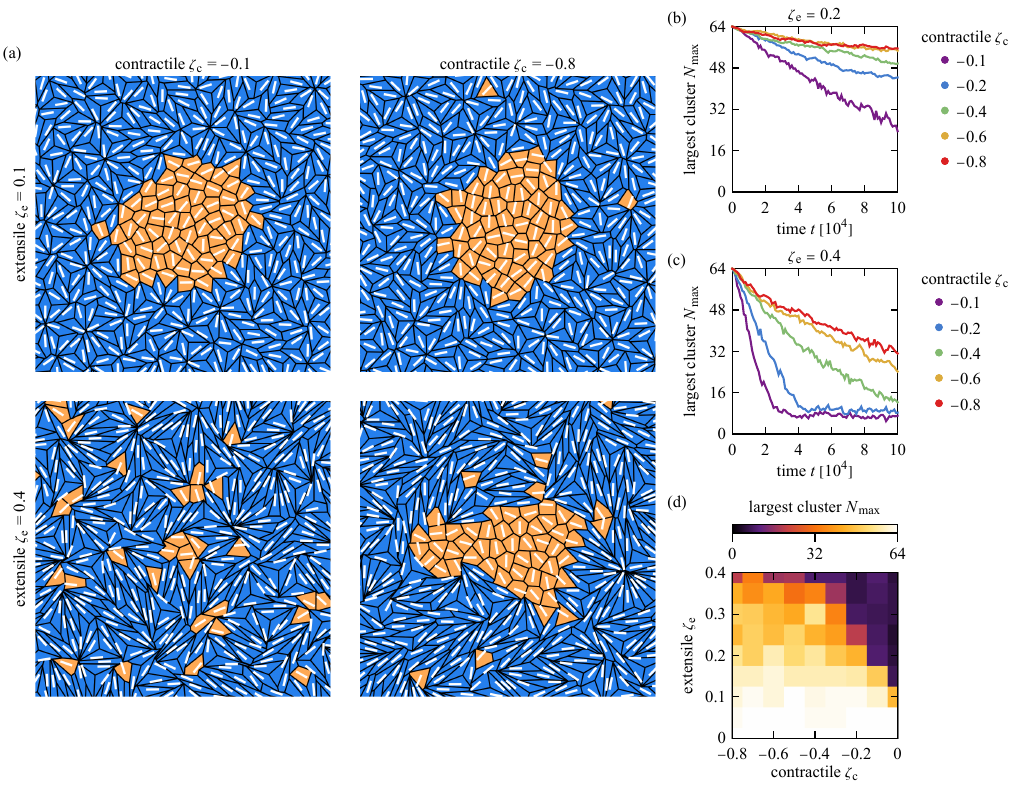}
    \caption{
    (a) Largest connected cluster of contractile cells at $t=5\times 10^4$ starting from a contractile (orange) cluster of 64 cells surrounded by an extensile (blue) bulk, for different combinations of extensile and contractile activity. (b),(c) Size of the largest contractile cluster for different contractile activities $\zc$ at $\ze=0.2$ (b) and $\ze=0.4$ (c); data averaged over 10 simulation runs at each set of parameters. (d) Size of the largest contractile cluster at $t=10^5$ as a function of extensile and contractile activity. Each value corresponds to one simulation at that set of parameters. $k_\text{P}=0.02$ and $p_0=3.95$ for all panels.}
    \label{fig:droplet}
\end{figure*}

Therefore, to further investigate why phase separation takes place, we simulate a small, 64 cell, contractile cluster in an extensile bulk until $t=10^5$ [Fig.~\ref{fig:droplet}(a)]. First selecting $\zc=-0.1$ for the cluster and $\ze=0.1$ for the bulk, we find that the majority of the contractile cells remain connected over time. However, if extensile activity is increased to $\ze=0.4$, the cluster no longer remains connected and its largest component at $t=5\times10^4$ only has a few cells. Conversely, keeping the higher extensile activity $\ze=0.4$ and increasing  the magnitude of the contractile activity to $\zc=-0.8$ improves the stability of the cluster. This shows that while a higher extensile activity aids in breaking the cluster apart, a higher magnitude of contractile activity holds the cluster together better, suggesting an effective attraction between contractile cells. This is also supported by Figs.~\ref{fig:droplet}(b),(c), which show how the average size of the largest cluster of connected contractile cells changes over time for different extensile and contractile activities, as well as Fig.~\ref{fig:droplet}(d), which shows the size of the largest cluster at the end of the simulation ($t=10^5$) for different $\ze$ and $\zc$. 

As evidence that the effective attraction is intrinsic to the contractile cells and does not depend on the presence of an extensile bulk, we simulate an alternative model in which the contractile cluster is surrounded by passive cells, but each contractile cell also experiences a polar force~\cite{sussman2017cellgpu} pointing in a constant random direction (see Supplemental Material, Sec.~\sectionpolar~\cite{SI} for details). Increasing contractile activity again improves the stability of the cluster (Fig.~\sipolar~\cite{SI}). A plausible reason for the effective attraction is that, as shown in Ref.~\cite{lin2023structure}, the contractile activity considered here increases the shear modulus of the tissue, rendering a vertex model solid-like even above the passive rigidity transition threshold. Therefore, a connected contractile region should oppose deformation and, as a consequence, resist being separated into smaller regions. To quantify the effective attraction, we measure the threshold force necessary to pull apart a pair of adjacent contractile cells in an otherwise passive tissue. This force increases approximately linearly with the magnitude of contractile activity. See Fig.~\sipair \ and Supplementary Material, Sec.~\sectionpair~\cite{SI} for details.

These observations suggest the following mechanism for phase separation: as in a purely extensile system, extensile activity induces chaotic flows that result in cell rearrangements. When contractile cells come into contact, they then tend to remain connected due to the contractile-activity-induced effective attraction. This tendency becomes more pronounced as the magnitude of contractile activity increases, whereas further increasing extensile activity hinders it, due to higher extensile activity inducing stronger chaotic flows which can then break apart connected contractile cells. 

If there is no extensile activity in the first place, flows do not develop, preventing the necessary rearrangements for cell sorting. Thus this also explains why contractile-contractile and contractile-passive mixtures do not phase separate [Figs.~\sisameactivity (a),(c)~\cite{SI}]. However, extensile-extensile and extensile-passive mixtures still show some phase separation, which indicates that mechanisms such as differential diffusivity discussed above also play a role [Figs.~\sisameactivity (b),(d)~\cite{SI}], noting that a higher activity in a purely extensile system corresponds to a higher diffusion coefficient [Fig.~\sisameactivity (e)~\cite{SI}]. 

For,  e.g., the parameter set shown in Fig.~\ref{fig:sorting}(b), a system that starts out fully phase separated has a higher passive energy after the start of the simulation when compared to an initially mixed system. Moreover, as the mixed system phase separates at that set of parameters, its passive energy increases (Fig.~\siinitial~\cite{SI}). This indicates that the phase separation is not primarily driven by the passive contributions of the model. Note, however, that there are also parameter combinations where sorting is accompanied by a reduction of the passive energy. Lastly, there is a small difference is cell areas between extensile and contractile cells [average area of $0.981$ and $1.019$, respectively, for the tissue in Fig.~\ref{fig:sorting}(b)]. While differences in cell area have been shown to lead to  clone dispersion and tissue fluidisation~\cite{ramanathan2019cell,bocanegra2023cell}, they do not appear to be driving sorting here: a five-fold increase in area elasticity reduces the difference in area so that the mean value is $0.996$ for extensile vs $1.004$ for contractile cells, but does not noticeably affect the extent of sorting (Fig.~\siarea \ and Supplemental Material, Sec.~\sectionarea~\cite{SI}). Moreover, it has been shown that differences in cell area do not lead to demixing on large scales in a vertex model~\cite{sahu2020sorting}.

{\textit{Discussion}}---We studied a mixture of extensile and contractile cells using an active nematic vertex model. For appropriate choices of activity, the cell mixture can fully separate, so that almost all extensile and almost all contractile cells each form a single contiguous region. More generally, the magnitude of the phase ordering at the end of the simulation depends monotonically on contractile activity with increasing magnitude of contractile activity leading to better (or equal) sorting. Conversely, the dependence on extensile activity is non-monotonic, with an optimal extensile activity magnitude for phase separation at a given set of parameters (Fig.~\ref{fig:sorting}). 

This behaviour can be interpreted by noting that extensile cells in the mixture are more motile than contractile ones~(Fig.~\ref{fig:msd}). They tend to stir the mixture so that cells undergo neighbour exchange. Contractile cells that come into contact as a result then tend to remain in a connected cluster due to contractile activity resulting in an effective attraction between those cells~(Fig.~\ref{fig:droplet}). If, however, the extensile cells are too active they are able to break up the clusters, hindering phase separation. Note that the cluster simulations shown in Fig.~\ref{fig:droplet} are also pertinent to understanding the conditions for a connected group of cells to separate, which is relevant to, e.g., clone dispersion~\cite{ramanathan2019cell,bocanegra2023cell} and metastasis~\cite{friedl2012classifying}. See Fig.~\sidispersion\ and Supplementary Material, Sec~\sectiondispersion~\cite{SI} for cluster dispersion analysis under different activity combinations.

This work contributes to the recent theoretical literature concerning extensile-contractile phase separation in active nematics~\cite{bhattacharyya2023phase,graham2024cell,bhattacharyya2024phase} in demonstrating strong sorting in the context of a model that resolves individual cells. The mechanism proposed here relies on effective attraction between contractile cells. It is, therefore, different from the phase separation in a mixture of two fluids with different nematic activities, which separate due to active anchoring at concentration gradients causing relative flows between the two fluids~\cite{bhattacharyya2023phase,bhattacharyya2024phase}. It would be interesting to analyse whether a continuum active nematic model extended to account for deformable nematogens~\cite{hadjifrangiskou2023active}, and therefore more closely resembling the active nematic vertex model, would lead to a similar phase separation mechanism as that discussed here.\\

\begin{acknowledgments}
{\textit{Acknowledgments}}---We wish to thank Matej~Krajnc for providing the initial version of the vertex model code and Saraswat Bhattacharyya, James N.~Graham, Ioannis Hadjifrangiskou, Chaithanya K.V.S., Francesco Mori, and Rastko~Sknepnek for many helpful discussions.  We acknowledge support from the UK Engineering and Physical Sciences Research Council (Award EP/W023849/1) and ERC Advanced Grant ActBio (funded as UKRI Frontier Research Grant EP/Y033981/1).
\end{acknowledgments}

\providecommand{\noopsort}[1]{}\providecommand{\singleletter}[1]{#1}%
\newpage

\end{document}


\newcommand{\zc}{\zeta_{\rm{c}}}
\newcommand{\ze}{\zeta_{\rm{e}}}

\newcommand{\figdroplet}{3}

\preprint{}

\title{
\textbf{Supplemental Material: \\ Cell Sorting in an Active Nematic Vertex Model}}
\author{Jan Rozman}
\email{jan.rozman@physics.ox.ac.uk}
\affiliation{Rudolf Peierls Centre for Theoretical Physics, University of Oxford, Oxford OX1 3PU, United Kingdom}
\author{Julia M. Yeomans}%
\affiliation{Rudolf Peierls Centre for Theoretical Physics, University of Oxford, Oxford OX1 3PU, United Kingdom}

\date{\today}

{
\let\clearpage\relax
\maketitle
}

\section{Active nematic forces}
To add active nematic forces to the vertex model as in Ref.~\cite{lin2023structure}, the stress tensor $\boldsymbol{\sigma}_i$ acting on a cell $i$ is translated into vertex forces following the approach proposed by Tlili \textit{et al.}~\cite{tlili2019shaping,lin2022implementation}. The force on a junction of cell $i$ between vertices $j$ and $j+1$ due to a stress tensor $\boldsymbol{\sigma}_i$ is 
\begin{equation}
    \mathbf{f}^\text{st}_{j,j+1}(\boldsymbol{\sigma}_i)=-l_{j,j+1} \boldsymbol{\sigma}_i\cdot\mathbf{n}_{j,j+1},
\end{equation}
where $l_{j,j+1}$ is the length of the junction and $\mathbf{n}_{j,j+1}$ is a unit vector normal to the junction and pointing out of the cell $i$. The force on vertex $j$ due to the stress $\boldsymbol{\sigma}_i$ is then taken to be half of the force due to that stress tensor on the junction between vertices $j$ and $j+1$ and half of the force on the junction between vertices $j-1$ and $j$:
\begin{equation}
    \mathbf{f}^\text{st}_{j}(\boldsymbol{\sigma}_i)=\frac{1}{2}\left[\mathbf{f}^\text{st}_{j,j+1}(\boldsymbol{\sigma}_i)+\mathbf{f}^\text{st}_{j-i,j}(\boldsymbol{\sigma}_i)\right].
\end{equation}

\section{T1 transitions}
To test how the results of this study depend on the T1 transition threshold, $\ell_\text{T1}$, and the length of the junction after the T1, $\ell'_\text{T1}$, we performed simulations corresponding to those used to generate Figs.~1(e),(f) and Fig. 2 of the main text for $\left(\ell_\text{T1},\ell'_\text{T1}\right)$ pairs $(0.01,0.01),\ (0.001, 0.0011),$ and $(0.001, 0.001)$ [the main text uses $(0.01,0.011)$]. Figure~\ref{sfig:threshold} shows the corresponding plots. The qualitative results remain similar for all four combinations and extensile-contractile phase separation always takes place for appropriate activities. However, the quantitative values of $SI$ and especially the cell motility do significantly depend on the choice of T1 thresholds. In particular, for the cases where $\ell_\text{T1}=\ell'_\text{T1}$, motion in the extensile-contractile mixture more often arrests over time at some lower activity magnitudes. This prevents further sorting and contributes to the difference in final $SI$ in that part of the parameter space when compared to the $\ell'_\text{T1}=1.1\times\ell_\text{T1}$ cases. An explanation for the qualitative dependence on $\left(\ell_\text{T1},\ell'_\text{T1}\right)$ values is that this model of extensile activity produces a high number of rosettes in which multiple vertices connected by short junctions lie near each other as discussed in Ref.~\cite{lin2023structure}; therefore, rosette resolution should depend finely on the details of T1 implementation.  Lastly, we note that T1 transitions are implemented such that  after the vertices are moved in each time step, every junction below the threshold length is first collapsed into a four-fold vertex.  Then all four-fold vertices are resolved into the final configuration. If a junction is connected to a four-fold vertex due to a neighbouring junction already undergoing a T1 transition on that same time step, it cannot itself also start a T1 transition.

\section{Active polar forces}\label{app:polar}
To add a polar force on each contractile cell for Fig.~\ref{sfig:polar}, we follow the approach proposed in Ref.~\cite{sussman2017cellgpu}. Each cell is assigned a polar activity $\alpha_i$ and an angle $\theta_i$. The additional polar active force on each vertex $j$ then reads
\begin{equation}
    \mathbf{f}^\text{polar}_{j}=\frac{1}{n_j}\sum_{i}^{n_j}\alpha_i \mathbf{n}_i,
\end{equation}
where the sum is over all $n_j$ cells that share vertex $j$ and 
\begin{equation}
    \mathbf{n}_i=(\cos(\theta_i),\sin(\theta_i))
\end{equation}
is a unit vector along the direction of the polarity of cell $i$. We set $\alpha_i=0$ for the passive cells and $\alpha_i=0.02$ for the contractile ones. At the start of the simulation, the angle $\theta_i$ for each cell is individually chosen from a random uniform distribution between $0$ and $2\pi$. Thereafter, the angles for all cells remain unchanged for the entire simulation.

\section{Quantifying effective contractile attraction}\label{app:pair}
To quantify the effective attraction between contractile cells, we preform simulations of two adjacent contractile cells surrounded by a passive tissue. A pair of polar forces pointing in opposite directions, implemented as in Sec.~\ref{app:polar}, act to pull the cells apart. Simulations run until $t=5\times 10^4$, and we scan over polar activities in increments of $0.01$ from $0$. Figure~\ref{sfig:pair}(a) shows the lowest polar activity $\alpha_*$ at which the two cells are separated by the end of the simulation for different $\zeta_c$. In this setup, the effective attraction can be explained as follows: as the polar forces pull the cells apart, they elongate in a direction perpendicular to their shared junction. Therefore, the contractile active force acts to extend that junction, as shown in Fig.~\ref{sfig:pair}(b).

\section{Cell area analysis}\label{app:area}
There is a small difference in area between extensile and contractile cells. Focusing on the model tissue shown in Fig.~1(b) of the main text, the mean and standard deviation of cell area for extensile cells are $0.981\pm0.008$ vs $1.019\pm0.007$ for contractile ones. Differences in cell area have been shown to fluidise a tissue and allow for clone dispersion~\cite{ramanathan2019cell,bocanegra2023cell}. To test if they play a major role in the phase separation observed here, we extend the passive energy function to include an area elasticity modulus $k_\text{A}$
\begin{equation}\label{eq:area}
    e_{\rm VM}=\sum_{i}\left[ \frac{k_\text{A}}{2}\left(a_i - 1\right)^2 + \frac{k_\text{P}}{2}\left(p_i - p_0\right)^2\right].
\end{equation}
The results reported elsewhere in the manuscript correspond to $k_\text{A}=1$. Increasing the area elasticity modulus to $k_\text{A}=5$ reduces the difference in area so that the mean and standard deviation are  $0.996\pm0.003$ for extensile vs $1.004\pm 0.003$ for contractile cells [Fig.~\ref{sfig:area}(a)]. However, the final segregation index $SI$ remains approximately the same in this range of $k_\text{A}$ [Fig.~\ref{sfig:area}(b)], suggesting that differences in cell area are not the main factor behind cell sorting. Moreover, it has been shown that differences in cell area in a vertex model do not lead to phase separation on large scales~\cite{sahu2020sorting}.

\section{Cluster dispersion}\label{app:dispersion}
Simulations of contractile clusters in an extensile bulk shown in Fig.~\figdroplet \ of the main text are also applicable to understanding the conditions under which an initially connected region of cells fragments, as in, e.g.,  clone dispersion~\cite{ramanathan2019cell,bocanegra2023cell} and metastasis~\cite{friedl2012classifying}. Here we, therefore, extend the analysis of the cluster system by also considering other combinations of cluster activity $\zeta_\text{cl}$ and bulk activity $\zeta_\text{bu}$ (Fig.~\ref{sfig:dispersion}). We find that an extensile cluster in an extensile bulk always disperses on the simulation timescale [Fig.~\ref{sfig:dispersion}(a)]. Inversely, a contractile cluster in a contractile bulk remains stable [Fig.~\ref{sfig:dispersion}(b)]. For extensile clusters in a contractile bulk, increasing the contractility of the bulk aids cluster stability, whereas increasing extensile activity aids dispersion [Figs.~\ref{sfig:dispersion}(c)-(e)].
\newpage

\section*{Supplementary figures}
\begin{figure}[h]
    \centering
    \includegraphics{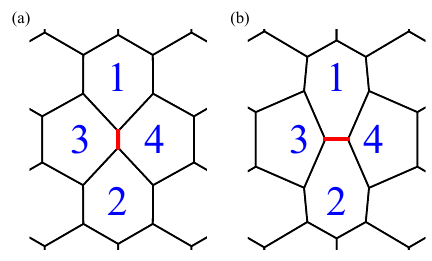}
    \caption{Schematic of a tissue before and after T1 transition: if the length of a junction falls below a threshold value $\ell_{T1}$, the topology of the tiling is modified so that the two previously unconnected cells [$1$ and $2$ on panel (a)] come into contact, whereas the two previously connected cells [$3$ and $4$ on panel (a)] are now separated. The final length of the new junction is set to $\ell'_\text{T1}=1.1\times \ell_\text{T1}$. For visibility, the junctions on both panels are shown longer than $\ell_\text{T1}$ and $\ell'_\text{T1}$, respectively.}
    \label{sfig:transition}
\end{figure}

\begin{figure}[h]
    \centering
    \includegraphics{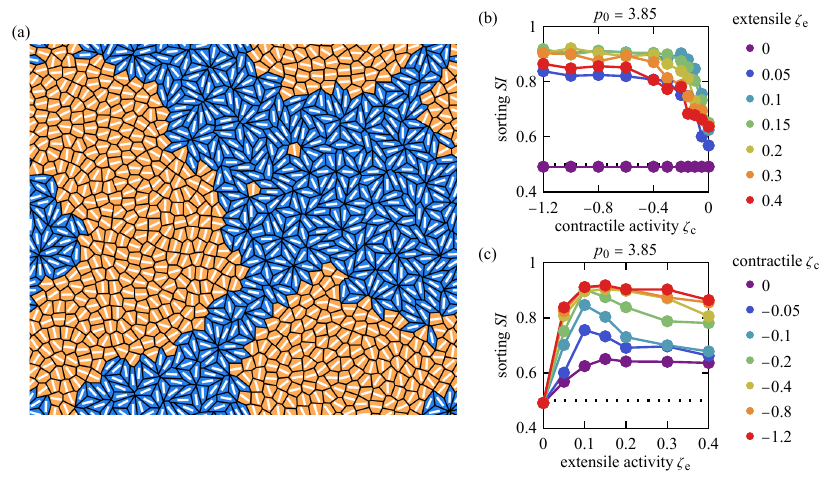}
    \caption{(a) Model tissue at $t=6\times10^5$ with $p_0=3.85$ for $\ze=0.1$ (blue) and $\zc=-0.4$ (orange); segregation index, $SI=0.90$.  (b) $SI$ at $t=6\times10^5$ for $p_0=3.85$ as a function of contractile activity at different extensile activities. (c) $SI$ at $t=6\times10^5$ for $p_0=3.85$ as a function of extensile activity at different contractile activities. Dotted lines on panels (b) and (c) show $SI=0.5$. $k_\text{P}=0.02$ for all panels.}
    \label{sfig:perimeter}
\end{figure}

\begin{figure}[h]
    \centering
    \includegraphics{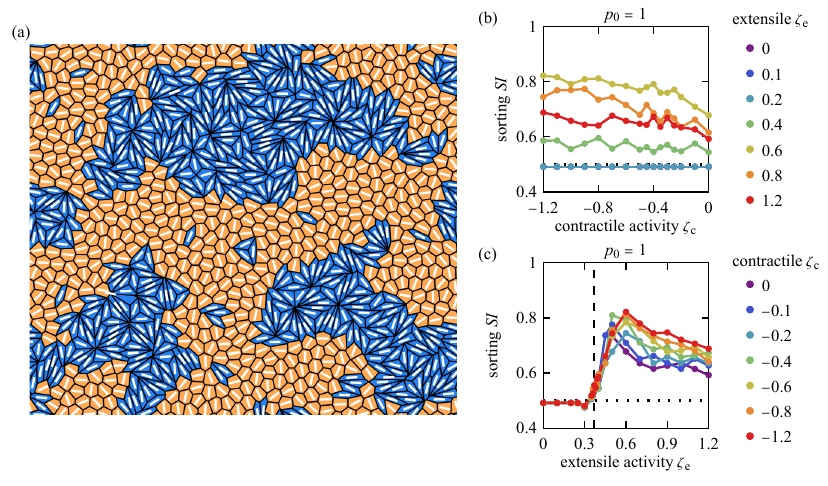}
    \caption{(a) Model tissue at $t=6\times10^5$ with $p_0=1$ (i.e., in the solid phase of the passive model) for $\ze=0.6$ (blue) and $\zc=-1.2$ (orange); segregation index, $SI=0.82$.  (b) $SI$ at $t=6\times10^5$ for $p_0=1$ as a function of contractile activity at different extensile activities. Values for $\ze=0$ and $\ze=0.1$ match those for $\ze=0.2$ and are therefore hidden under them. (c) $SI$ at $t=6\times10^5$ for $p_0=1$ as a function of extensile activity at different contractile activities. Dotted lines on panels (b) and (c) show $SI=0.5$. Dashed line on panel (c) shows $\ze=0.37$, the reported threshold activity for tissue fluidisation at $p_0=1$~\cite{lin2023structure}. $k_\text{P}=0.02$ for all panels.}
    \label{sfig:solid}
\end{figure}

\begin{figure}[h]
    \centering
    \includegraphics{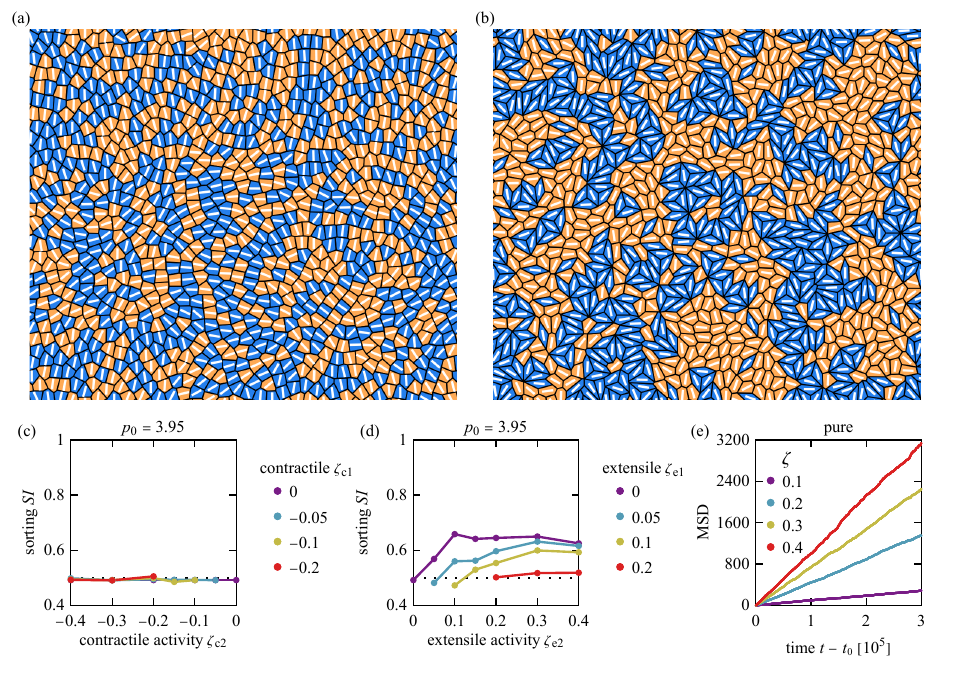}
    \caption{(a) Model tissue at $t=6\times10^5$: orange cells have $\zeta_i=\zeta_\text{c1}=-0.4$ and blue cells have $\zeta_i=\zeta_\text{c2}=-0.1$; $SI=0.49$.  (b)~Model tissue at $t=6\times10^5$: blue cells have $\zeta_i=\zeta_\text{e1}=0.1$ and orange cells have $\zeta_i=\zeta_\text{e2}=0$; $SI=0.66$. (c) Segregation index at $t=6\times10^5$ for different activities in a mixture of contractile-contractile or contractile-passive cells. (d) Segregation index at $t=6\times10^5$ for different activities in a mixture of extensile-extensile or extensile-passive cells. Dotted lines on panels (c) and (d) show $SI=0.5$. (e) Mean-squared displacement in a pure extensile system for different extensile activities. $k_\text{P}=0.02$ and $p_0=3.95$ for all panels.}
    \label{sfig:excont}
\end{figure}

\begin{figure*}
    \includegraphics{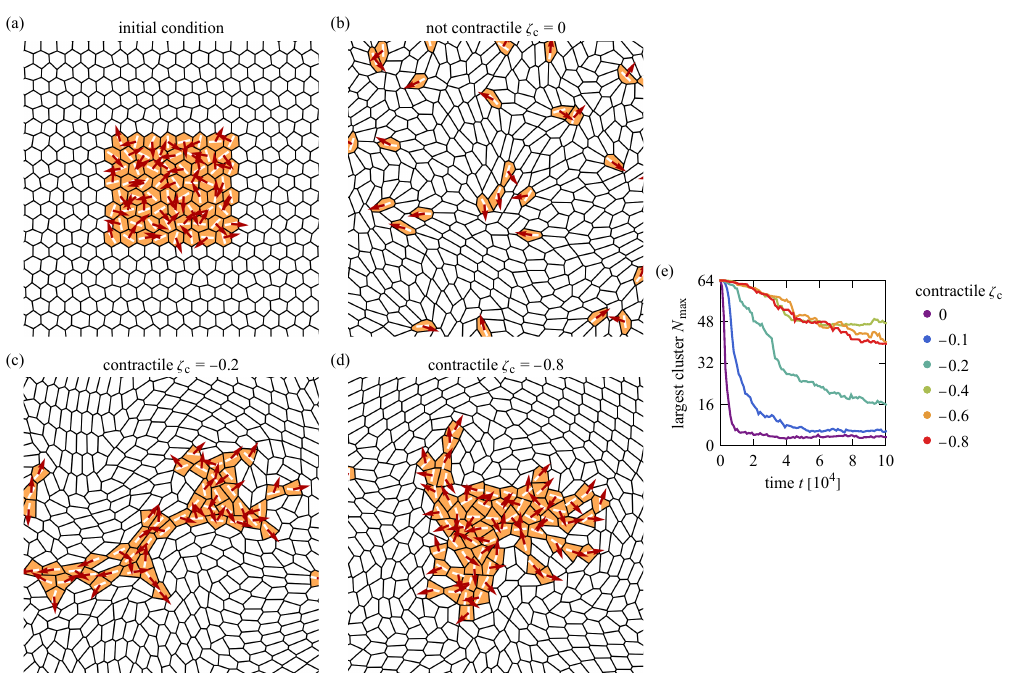}
    \caption{
    (a) Initial conditions for simulations with a cluster of 64 contractile and polar cells (orange) surrounded by passive cells (white). Red arrows show the direction of the polar force and white lines show cell directors. (b-d) Resulting largest cluster of connected polar and contractile cells at $t=2.5\times 10^4$ for no contractile activity (b), $\zc=-0.2$ (c), and~$\zc=-0.8$ (d). (e) Size of the largest cluster of connected polar and contractile cells as a function of time for different contractile activities, averaged over 10 simulation runs. $k_\text{P}=0.02$ and $p_0=3.95$ for all panels.}
    \label{sfig:polar}
\end{figure*}

\begin{figure}[h]
    \centering
    \includegraphics{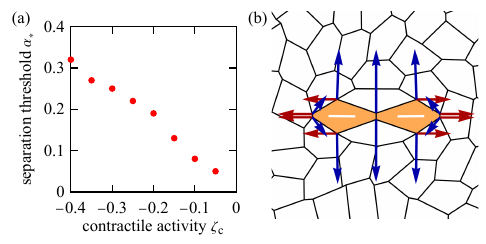}
    \caption{(a) Threshold polar activity necessary for a pair of polar forces pointing in opposite directions to separate two initially adjacent contractile cells by $t=5\times 10^4$ in an otherwise passive tissue as a function of contractile activity. (b) Pair of contractile cells being pulled in opposite directions by polar forces at $t=5\times 10^4$, with $\alpha=0.1$ and $\zeta_c=-0.2$. Orange cells are contractile and polar, white cells are passive, white lines show cell directors, red arrows show polar forces on vertices, and blue arrows show contractile forces on vertices. $k_\text{P}=0.02$ and $p_0=3.95$ for both panels.}
    \label{sfig:pair}
\end{figure}

\begin{figure*}
    \includegraphics{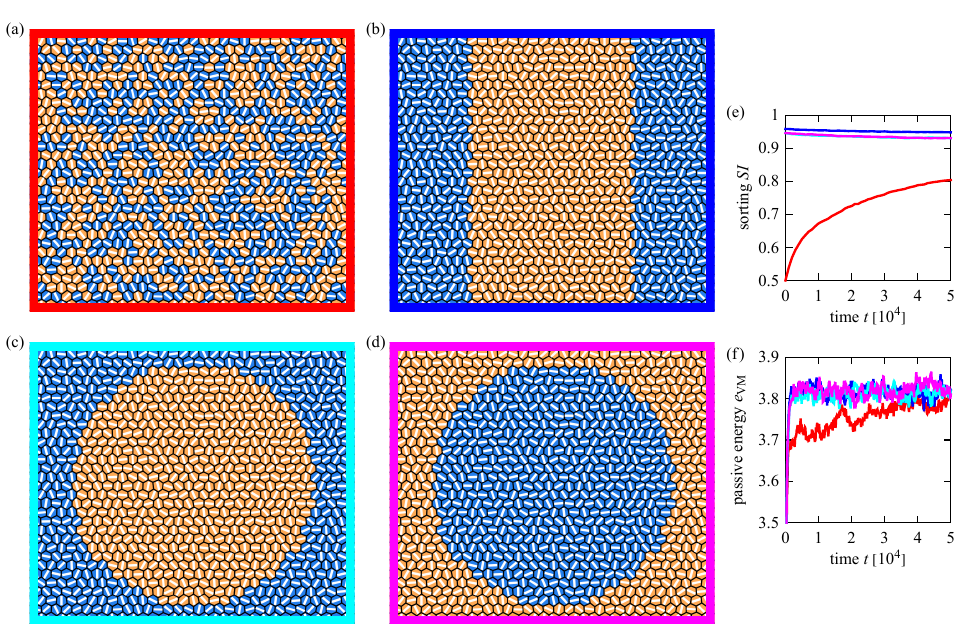}
    \caption{
    (a)-(d) Four different initial conditions for the simulation: random mixture of extensile and contractile cells (a), parallel stripes of contractile and extensile cells (b), a disk of contractile cells surrounded by extensile cells (c), and a disk of extensile cells surrounded by contractile cells (d). All four have $512$ contractile and $512$ extensile cells. The last two initial conditions are included in case the energy depends on the curvature of the boundary. (e) Segregation index $SI$ as a function of time for the four types of initial conditions, with the colour of the plots matching the border of the corresponding initial conditions on panels (a)-(d). (f) Passive energy $e_\text{VM}$ as a function of time for the four types of initial conditions, with the colours again matching the ones of panels (a)-(d). Panels (e) and (f) are averaged over $20$ simulations starting from each type of initial condition (the starting perturbation and, in the mixed case, selection of extensile and contractile cells differ between runs, leading to different trajectories); $\ze=0.1$, $\zc=-0.4$,  $k_\text{P}=0.02$, and $p_0=3.95$.
    }
    \label{sfig:initial}
\end{figure*}

\begin{figure}[h]
    \centering
    \includegraphics{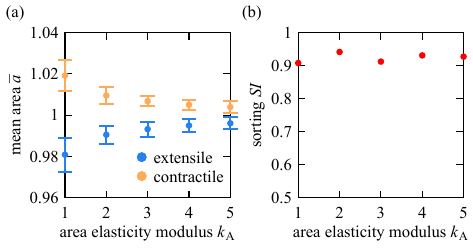}
    \caption{(a) Mean area of extensile and contractile cells at $t=6\times 10^5$ for $\ze=0.1$ and $\zc=-0.4$ as a function of the area elasticity modulus $k_\text{A}$ [Eq.~\eqref{eq:area}]. (b) Segregation index $SI$ at $t=6\times10^5$ for $\ze=0.1$ and $\zc=-0.4$ as a function of the area elasticity modulus $k_\text{A}$. $k_\text{P}=0.02$ and $p_0=3.95$ for both panels.}
    \label{sfig:area}
\end{figure}

\begin{figure}[h]
    \centering
    \includegraphics{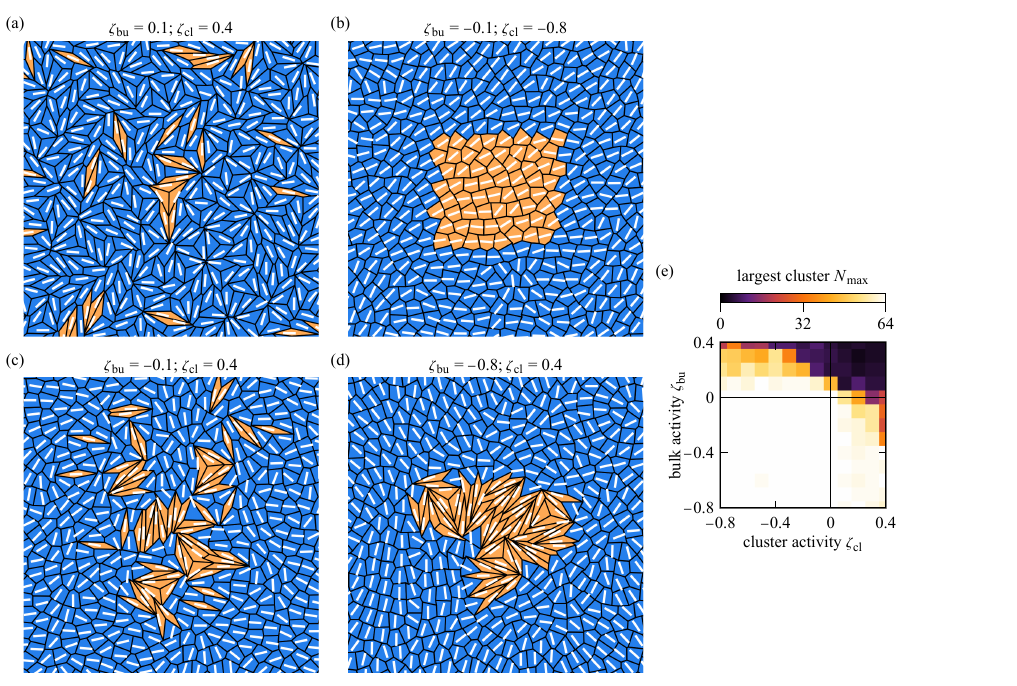}
    \caption{(a)-(d) Largest cluster of connected  cells with activity $\zeta_\text{cl}$ (orange) at $t=5\times 10^4$, starting from a cluster of 64 cells surrounded by a bulk with cell activity $\zeta_\text{bu}$ (blue), for: $(\zeta_\text{bu}=0.1, \zeta_\text{cl}=0.4)$ (a), $(\zeta_\text{bu}=-0.1, \zeta_\text{cl}=-0.8)$ (b), $(\zeta_\text{bu}=-0.1, \zeta_\text{cl}=0.4)$ (c), and $(\zeta_\text{bu}=-0.8, \zeta_\text{cl}=0.4)$ (d). (e) Size of the largest cluster at $t=10^5$ as a function of cluster activity $\zeta_\text{cl}$ and bulk activity $\zeta_\text{bu}$. Each value corresponds to one simulation at that set of parameters. $k_\text{P}=0.02$ and $p_0=3.95$ for all panels.}
    \label{sfig:dispersion}
\end{figure}

\begin{figure*}
    \includegraphics{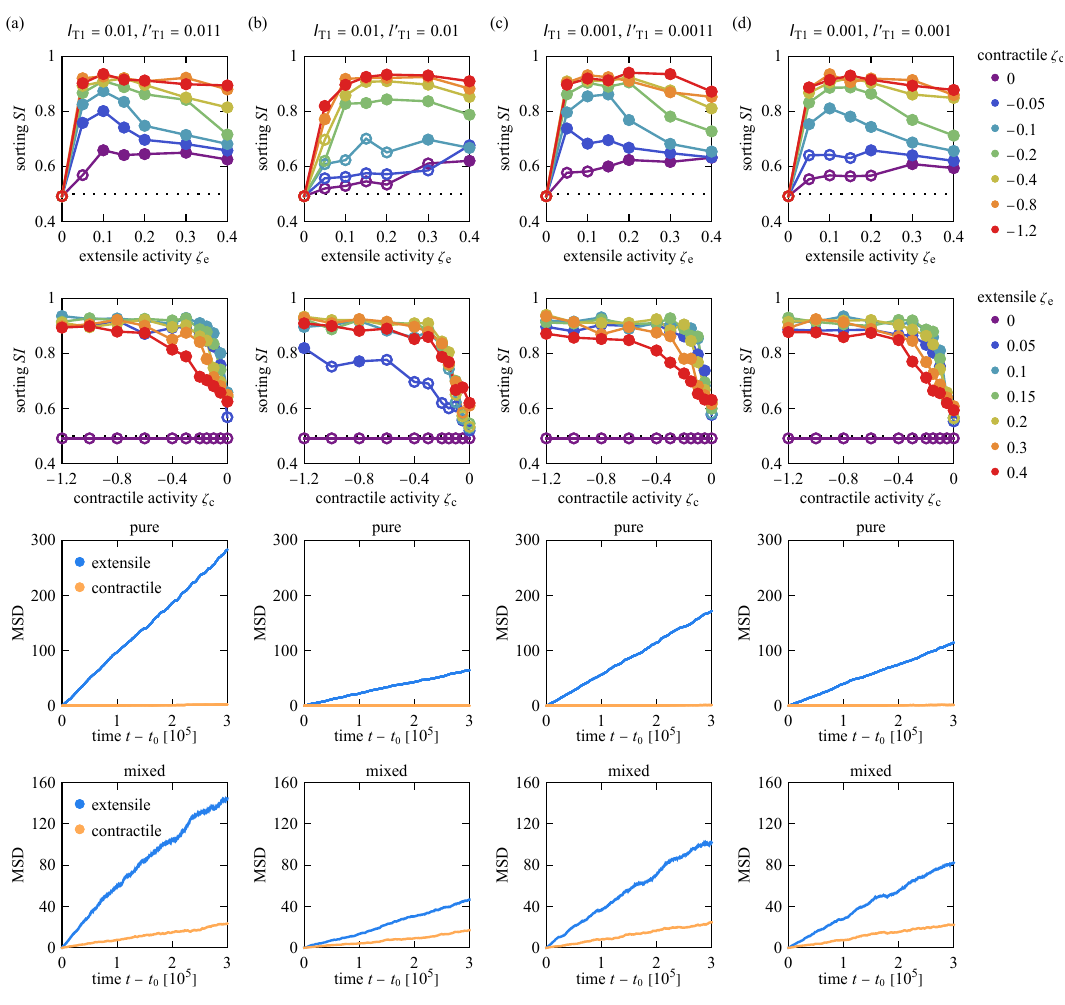}
    \caption{
    (a)-(d) From top to bottom: segregation index at $t=6\times10^5$ as a function of extensile activity at different contractile activities, segregation index at $t=6\times10^5$ as a function of contractile activity at different extensile activities, MSD of a system of only extensile ($\ze=0.1$) or only contractile ($\zc=-0.4$) cells, and MSD of a 50:50 mixture of extensile ($\ze=0.1$) and contractile ($\zc=-0.4$) cells, for $\left(\ell_\text{T1}=0.01,\ \ell'_\text{T1}=0.011\right)$ (a; same parameters as main text), $\left(\ell_\text{T1}=0.01,\ \ell'_\text{T1}=0.01\right)$ (b), $\left(\ell_\text{T1}=0.001,\ \ell'_\text{T1}=0.0011\right)$ (c), and $\left(\ell_\text{T1}=0.001,\ \ell'_\text{T1}=0.001\right)$ (d). Full circles in the segregation index plots correspond to parameter sets where the change in MSD between $t=4\times10^5$ and $t=6\times10^5$ is greater than 0.001, whereas empty circles correspond to parameter sets where it is below that threshold, indicating motion has arrested. $k_\text{P}=0.02$ and $p_0=3.95$ for all panels.
    }
    \label{sfig:threshold}
\end{figure*}

\clearpage
\newpage

\section*{Movie captions}
	\noindent Movie S1. Model tissue evolution of an extensile-contractile mixture showing cell sorting. $k_\text{P}=0.02$, $p_0 = 3.95$, $\ze = 0.1$, and $\zc = -0.4$; simulation runs until $t=6\times10^5$.\\ 

\bibliographystyle{apsrev4-1}
\providecommand{\noopsort}[1]{}\providecommand{\singleletter}[1]{#1}%